\documentclass[aps,pre,twocolumn,floatfix,nofootinbib,amsmath,amssymb]{revtex4}

\usepackage[english]{babel}
\usepackage{graphicx}
\usepackage{textcomp}
\usepackage{amsmath}
\usepackage[usenames,dvipsnames]{xcolor}
\def\barp{p}
\def\barq{q}

\begin{document}

\title{Exact probability distribution functions for Parrondo's games}

\author{Rubina Zadourian$^1$}
\author{David B. Saakian$^{2,3}$}
\email{saakian@yerphi.am} \author{Andreas Kl\"umper$^4$}
\affiliation{$^1$Max Planck Institute for the Physics of Complex Systems, Dresden, Germany  }
\date{\today}

\affiliation{$^2$Institute of Physics, Academia Sinica, Nankang,
Taipei 11529, Taiwan}

\affiliation{$^3$ A.I. Alikhanyan National Science Laboratory
(Yerevan Physics
Institute) Foundation,\\
 2 Alikhanian Brothers St., Yerevan 375036, Armenia}
\affiliation{$^4$ Wuppertal University, Germany}

\begin{abstract}
We consider discrete time Brownian ratchet models: Parrondo's
games. Using the Fourier transform, we calculate the exact
probability distribution functions for both the capital dependent
and history dependent Parrondo's games. We find that in some cases
there are oscillations near the maximum of the probability
distribution, and after many rounds there are two limiting
distributions, for the odd and even total number of rounds of
gambling. We assume that the solution of the aforementioned models
can be applied to portfolio optimization.

\end{abstract}

\maketitle


The Parrondo's games \cite{pa96}-\cite{ab10}, related to the Brownian ratchets
\cite{pr92}-\cite{ch14} are interesting phenomena at the intersection of game
theory, econophysics and statistical physics, see \cite{ab10} for the
inter-disiplinary applications. In case of Brownian ratchets the particle
moves in a potential, which randomly changes between 2 versions. For each
there is a detailed balance condition. On average there is a motion due to the
random switches between two potentials. The phenomenon is certainly related to
portfolio optimization \cite{so11},\cite{so15}. In the related situation in
economics, one is using the ``volatility pumping'' strategy in portfolio
optimization, for two asset portfolios, keeping one half of the capital in the
first asset, the other half in the second asset with high volatility
\cite{vo}.

J.~M.~R.~Parrondo invented his game following Brownian ratchets for the
discrete time case \cite{pa96}, to model gambling. An agent uses two biased
coins for the gambling, and both strategies are loosing. In some cases a
random combination of two loosing games is a winning game. It is interesting
that the opposite situation is also possible, a random combination of two
winning games can give a loosing game.

Parrondo's games have been considered either on the one dimensional axis with
some periodic potential, or by looking at the time dependent version of the
game parameters.

For the first case the state of the system is characterized by the current
value of the money $X$, and the choice of the strategy.  There is a period $M$
defining the rules, how capital $X$ can move up or down. The rules of the game
depend on Mod$(X,M)$, where $M$ is the period of the ``potential''.
Originally $M=3$ games were considered, then $M=2$ versions of Parrondo's
games were constructed \cite{ab02}, \cite{as05}. For the history dependent
versions of the game the current rules of the game depend on the past, whether
there was a growth of capital in the previous rounds or not.

Later many modifications of the games were invented, i.e.~different integers
$M$ for both games \cite{sz14}, the Allison mixture \cite{ab14}, where the
random mixing of two random sequences creates some autocorrelation
\cite{ad09a}, and two envelope game problems \cite{ad10a}.  Especially
intriguing is the recent finding of a Parrondo's effect-like phenomenon in a
Bayesian approach to the modelling of the work of a jury \cite{gu16}: the
unanimous decision of its members has a low confidence. All the mentioned
works consider the situation with random walks, when there are random choices
between different strategies during any step, yielding a qualitatively
different result than in case of a fixed strategy. In \cite{gu16} we have a
single choice between the strategies during all the rounds.

In our recent work \cite{sa16} we calculated the variance for the history
independent case, as the variance of the distributions (volatilities) is
important in economics. Now we apply a Fourier transform technique to solve
exactly the probability distribution function.  The complete distributions are
useful to obtain the unknown parameters of the models, describing the data.
We apply this method to the random walks on a strip of the chains \cite{sa12},
which is the case of the capital dependent Parrondo's model. We calculate
the entire probability distribution for the capital, then solve the same
problem for the history dependent game case.\\

{\bf A biased discrete space and time random walk}.

As an illustration let us consider the discrete time random walk on the 1-$d$
axis, where the probability of right and left jumps are $\barp$ and $q$,
respectively. We can write the master equation for probability $P(n,t)$
at position $n$ after $t$ steps:
\begin{eqnarray}
&&P(n,t+1)=\label{e1}\\
&&\barp P(n-1,t)+q P(n+1,t)+(1-\barp-q)P(n,t).\nonumber
\end{eqnarray}
We have $P(n,0)=\delta_{n,0}$.

For the motion on the infinite axis we can always write a Fourier
transform
\begin{eqnarray}
\label{e2} P(n,t)&=&\int_{-\pi}^{\pi}dk e^{ikn}\bar
P(k,t),\nonumber\\
\bar P(k,t)&=&\frac{1}{2 \pi}\sum_n P(n,t)e^{-ikn}
\end{eqnarray}
Let the particle start at $n=0$, thus $\bar P(k,0)={1}/{2\pi}$.

Eq. (1) transforms into
\begin{eqnarray}
\label{e3} \bar P(k,t+1)= \left[pe^{-ik}+qe^{ik}+(1-p-q)\right]\bar P(k,t).
\end{eqnarray}
We obtain a solution
\begin{eqnarray}
\label{e4} \bar P(k,t)=\lambda^t(k)\cdot\bar P(k,0)\nonumber\\
\lambda(k)= \left[pe^{-ik}+qe^{ik}+(1-p-q)\right]
\end{eqnarray}
As $\lambda$ is a linear polynomial in $e^{ik}$ and $e^{-ik}$,
$\lambda^t$ is a polynomial with powers from $e^{-tik}$ to
$e^{tik}$.

We can write the solution as
\begin{eqnarray}
\label{e5} P(n,t)=
\int_{-\pi}^{\pi} dk e^{tV(ik)+ikn} \bar P(k,0),\nonumber\\
 V(\kappa):=\ln \left[pe^{-\kappa}+qe^{\kappa}+(1-p-q)\right],
\end{eqnarray}
where we defined the function $V(\kappa)$.
Using that $\lambda^t$ has a finite expansion in
powers of $e^{ik}$ we obtain for $\bar P(k,0)={1}/{2\pi}$
\begin{eqnarray}
\label{e6} P(n,t)=\frac{1}{2 t}\sum_{m=1}^{2t}
e^{tV(im\pi/t)+imn\pi/t}
\end{eqnarray}
While looking at more involved models, we will use both
presentations Eqs. (5),(6).

Note that eqs. (5) and (6) are exact for any $t$ and $n$. By use of the
saddle point approximation we derive the large $t$ and $n=xt$ asymptotics.
We are allowed to move the integration contour because of the analytic
dependence of the integrand on $k$ resulting in ($\kappa=i k$)
\begin{eqnarray}
\label{e7} P(n,t)=\frac{\exp[tu(x)]}{\sqrt{2 \pi t V''(\kappa)}},
\nonumber\\ x=-V'(\kappa),u(x)=V(\kappa)+\kappa x
\end{eqnarray}
As $u(x)$ is the Legendre transform of $-V(\kappa)$, we also have
$V''(k)=-1/u''(x)$ and hence
\begin{equation*}
P(n,t)=\frac{1}{\sqrt{2 \pi t}}\exp\left[tu(x)+1/2\log|u''(x)|\right],
\end{equation*}

Let us assume an expansion for $V(\kappa)$
\begin{eqnarray}
\label{e8}V(\kappa)=r\kappa+K \kappa^2/2
\end{eqnarray}
It then follows that
\begin{eqnarray}
\label{e9} <n>=rt,\nonumber\\
<(n-<n>)^2>=Kt
\end{eqnarray}

{\bf The case of several chains}


Consider the case of a random walk on the one dimensional axis,
using some rules  with period $M$.

We divide the entire $x$ axis in intervals of length $M$,
$[(n-1)M,nM],n\ge 1$, considering $l=\rm{Mod} (X,M)$. We represent
the  sets of $p_X$ (the discrete probability distribution of the
capital value) as $P_l(n,t), 0\le l< M$. An integer $t$ represents
the time.

One can write the following master equation \cite{mo00}
\begin{eqnarray}
\label{e10}
P_l(n,t+1)&=&A_{l_-}P_{l_-}(\hat n,t)+B_{l_+}P_{l_+}(\bar
n,t)\nonumber\\ &&+(1-A_{l}-B_{l})P_{l}(n,t)
\end{eqnarray}
 where  $l_-={\rm Mod}(l-1,M),  {l_+={\rm Mod}(l+1,M)}$,   $\hat n=n$ for $l-1\ge 0$ and $\hat n=n-1$ for $l-1< 0$;  $\bar
n=n$ for $l+1<M$ and $\bar n=n+1$ for $l+1> M$.  Thus the model is
characterized by the parameters $A_l,B_l$,  where $A_l$ and $B_l$
are the probabilities to win and lose for the capital $X$ with
$l=\rm {Mod} (X,M)$.

We again consider the Fourier transform
\begin{eqnarray}
\label{e11}
P_l(n,t)=
\int_{-\pi}^{\pi}dk e^{ikn}\bar P_l(k,t)
 \end{eqnarray}
Then we obtain
 \begin{eqnarray}
 \label{e12}
\bar P_l(k,t+1)&=&A_{l_{-}}e^{i k(\hat n-n)}\bar P_{l_{-}}(k,t)
+B_{l_+}e^{i k(\bar n-n)}v_{l_+}\nonumber\\&&+
(1-(A_{l}+B_{l}))\bar P_l(k,t)\nonumber\\
&\equiv& \sum_{m=0}^{M-1} \hat Q_{lm}(ik)\bar P_m(k,t)
 \end{eqnarray}
Using the eigenvalues and eigenvectors $\lambda_m(\kappa),v_{ml}(\kappa)$
($m=0,...,M-1$), of the matrix $Q(\kappa)$,
we find
\begin{eqnarray}
\label{e13} \bar P_l(n,t)=
\int_{-\pi}^{\pi}dke^{ikn}\sum_mc_m\exp[tV_m]v_{ml},
\end{eqnarray}
where $V_m(\kappa):=\ln (\lambda_m(\kappa))$ and $c_m(\kappa)$ are factors
determined by the initial distribution.
While using Eq.~(7), we choose as $V(\kappa)$ the eigenvalue
function $V_m(\kappa)$ yielding the maximal value of $u(x)$. Close
to the maximum of the distribution of $u(x)$ there is a single
choice, later we see a possibility for different sub-phases in our
model, related to the existence of M different eigenvalues. The
large deviation (decoding error) probability in optimal coding
theory \cite{gal} is similar to our case and has several
sub-phases.  To compare our results for the rate with the formulas
in
  \cite{ab02} we have to multiply the rate in Eq.~(9) with a factor $M$, as
  one step in $n$ in our approach equals $M$ ordinary steps.
 We deduced Eqs. (8),(9) for the mean growth of capital, i.e. for
 the case of the random walk on the strip.

{\bf The eigenvalues $\pm 1$}

Next we investigate more closely the case $A_{l}+B_{l}=1$ for all $l$ in
(\ref{e12}).

Consider first the case of odd $M$: $Q(0)$ has one eigenvalue +1 with left
eigenstate $(1,1,1,1...)$, and $Q(\pi i)$ has one eigenvalue -1 with left
eigenstate $(1,-1,1,-1...)$ and the two ``wavenumbers'' $\kappa=0$ and
$\kappa=\pi i$ contribute in the Fourier representation to the stationary
state.

Let us now consider the matrices $Q(\kappa)$ and $Q(\kappa+\pi i)$ for
arbitrary $\kappa$. It is easy to see that the spectra are simply related. Let
$(x_0,x_1,x_2,...)^T$ be a right eigenvector of $Q(\kappa)$ with eigenvalue
$\lambda(\kappa)$, then $(x_0,-x_1,x_2,...)^T$ is a right eigenvector of
$Q(\kappa+\pi i)$ with eigenvalue $-\lambda(\kappa)$.

Let as assume an expansion like Eq.~(8) for the leading $V(\kappa)$ near
$\kappa=0$, then
\begin{eqnarray}
\label{e14}V(\pi +\kappa)=\pi i+ r\kappa+K \kappa^2/2
\end{eqnarray}
Let $v^+$ and $v^-$ the right eigenstates of  $Q(0)$ and $Q(\pi i)$ with
eigenvalues $+1$ and $-1$. There are constants $\alpha$ and $\beta$ such that
%
%
%
%
\begin{eqnarray}
\label{e16} P_l(n,t)=\frac{\alpha v_l^++(-1)^{n+t}\beta v^-_l}{\sqrt{2Kt
\pi}}e^{-\frac{(n-rt)^2}{2Kt}}
\end{eqnarray}

We see oscillations caused by the rapid sign change of the second term. The
coefficients $\alpha$ and $\beta$ are determined by the initial probability
distribution. If this has been concentrated in $n=l=0$ then $\alpha=\beta$
(with $v^+$ and $v^-$ related as pointed out above). In this case $P_l(n,t)$
is non-zero (zero) for even (odd) $l+n+t$.


Consider now the case of even $M$. Here, $\hat Q(0)$ has one eigenvalue 1 with
left eigenstate $(1,1,1,1...)$ and one eigenvalue -1 with left eigenstate
$(1,-1,1,-1...)$. Hence, only the ``wavenumber'' $\kappa=0$ contributes in the
Fourier representation to the stationary state, but with two eigenvalues.  Let
$(x_0,x_1,x_2,...)^T$ be a right eigenvector of $Q(\kappa)$ with eigenvalue
$\lambda(\kappa)$, then $(x_0,-x_1,x_2,...)^T$ is also a right eigenvector of
$Q(\kappa)$, but with eigenvalue $-\lambda(\kappa)$. Now we find similar to
the case above
\begin{eqnarray}
\label{e16} P_l(n,t)=\frac{\alpha v_l^++(-1)^{t}\beta v^-_l}{\sqrt{2Kt
\pi}}e^{-\frac{(n-rt)^2}{2Kt}}.
\end{eqnarray}
Note that the oscillating factor does not depend on $n$. For a probability
initially concentrated in $n=l=0$ we find $P_l(n,t)$ is non-zero (zero) for
even (odd) $l+t$.

The findings in both cases, odd $M$ and even $M$, can however be summarized:
$P_l(n,t)$ is non-zero (zero) for even (odd) $l+n M+t$.

It is quite interesting to consider the quantity
\begin{eqnarray}
\label{e18} \hat P(n,t)=\sum_{l=0}^{M-1}P_l(n,t)
\end{eqnarray}
It shows non-zero oscillations in dependence on $n$ and $t$ for odd
$M$. Such oscillations do not exist for even $M$. The reason for this is
easily understood: $(1,-1,1,-1...)$ and $(x_0,x_1,x_2,...)^T$ are left and
right eigenvector of $Q(0)$ with eigenvalues $-1$ and $+1$. Hence their
product must be zero. This product, however, is equal to the term entering
$P(n,t)$ with a $(-1)^t$ factor.

{\bf Two expressions for the capital growth rates}
Consider capital depending Parrondo's game with $p_1,...,p_M$
for the winning probabilities for the corresponding Mod$(X,M)$.
We have a corresponding Matrix
\begin{equation}
Q(\kappa)=\begin{pmatrix}
0&p_1&...&p_{M}e^{-\kappa}\\
.&.&.&.\\
.&.&.&.\\
q_1e^\kappa&p_1&...&0
\end{pmatrix}
\end{equation}
Applying the method of \cite{ab02} gives
 \begin{eqnarray}
 \label{c1}
r=\frac{\sum_i(p_i-q_i)x_i}{\sum_ix_i}
 \end{eqnarray}
We prove that Eqs.~(\ref{e8},\ref{e9}) give the same result.  Let us denote by
$\lambda(\kappa)$ the largest eigenvalue of $Q(\kappa)$ with left and right
eigenstates $\langle y(\kappa)|$ and $|x(\kappa)\rangle$. For $\kappa=0$ we
have $\lambda(0)=1$ and $\vec y=(1,1,...,1)$. The growth rate $r$ is the first
derivative of of $\log\lambda(\kappa)$ at $\kappa=0$. As $\lambda(0)=1$ we
find $r=\lambda'(0)$, hence
 \begin{eqnarray}
 \label{c2}
r=\frac\partial{\partial\kappa}\frac{\langle y(\kappa)|
  Q(\kappa)|x(\kappa)\rangle}{\langle y(\kappa)|x(\kappa)\rangle}
=\frac{\langle y(0)|Q'(0)|x(0)\rangle}{\langle y(0)|x(0)\rangle}
 \end{eqnarray}
where the last equality follows from the Hellmann-Feynman theorem. Using the
explicit form of the matrix $Q(\kappa)$, $\langle y(0)|=(1,1,...,1)$, and
$|x(0)\rangle=(x_1,x_2,...,x_M)^T$ we find
\begin{equation}
\label{c3}
r=\frac{p_Mx_M-q_1x_1}{\sum_ix_i}
\end{equation}

Now we prove the equivalence of Eq.~(\ref{c1}) and Eq.~(\ref{c3}). The
eigenvalue equation for the right eigenstate $(x_1,x_2,...,x_M)^T$ of Q(0) for
eigenvalue 1 is
\begin{equation}
 \label{c4}
p_{i-1}x_{i-1}+q_{i+1}x_{i+1}=x_i
\end{equation}
for all $i$. From this we derive
\begin{equation*}
 \label{c5}
p_{i-1}x_{i-1}-q_ix_i = x_i-q_{i+1}x_{i+1}-q_ix_i=
p_ix_i-q_{i+1}x_{i+1}
\end{equation*}
where the first equality is simply (\ref{c4}) and the second equality is due to
$q_{i+1}=1-p_{i+1}$. Hence $p_{i-1}x_{i-1}-q_ix_i$ is independent of $i$
and the sum over this term for all $i$ is simply $M$ times the first
term for $i=1$. The sum over all terms can be written like
\begin{equation}
\sum_i(p_i-q_i)x_i=M(p_0x_0-q_1x_1) = M(p_Mx_M-q_1x_1)
\end{equation}
where we used cyclic ``boundary condition'' $x_0=x_M$.
This completes the proof.

{\bf The M=3 Parrondo's games}.
 Let us apply the theory of the
previous subsection to the concrete case of $M=3$ Parrondo's game.
We have two games. The first game is a random walk on the 1-d axis
with probability  $h$ for the right jumps and probability $(1-h)$
for the left jumps. For the second game the jump parameters depend
on the capital value.  The probability for the right jumps is  $r$
for mod$(X,3)\ne 0$ and $s$ for the case mod$(X,3)= 0$. We
randomly choose the game every round.

We solve the master equation (10) for calculating the probability distribution
after $t$ rounds. The results of iterative numerics are given in Figures (1),
(2).

We see that after $t=50$ there is an oscillation near the maximum,
then as time passes the number of oscillations grows.

\begin{figure}
\centerline{\includegraphics[width=.4\textwidth]{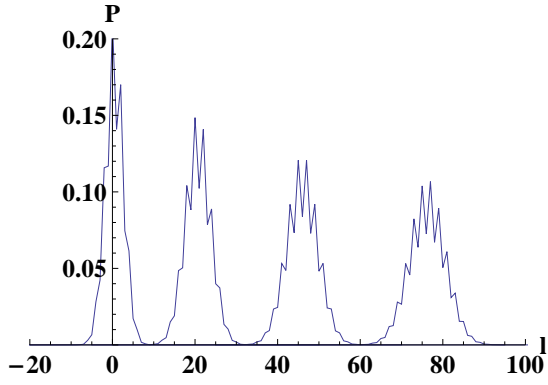}}
\caption{The  probability distribution for the capital growth
$\hat P(n,t)$ (see Eq.~(17)) for $t=50,150,200$ for the $M=3$
Parrondo's model with the
$p=0.5-ep,p_1=0.75-ep,p_2=0.1-ep,ep=0.005$, see Eq. (27). We moved
the distributions horizontally for the proper illustration. Our
analytical results by Eq.(6) are identical to the results of
numerics.} \label{fig2}
\end{figure}

 Let us
derive this distribution analytically.

We obtain for the matrix
$\hat Q(\kappa)$
 \begin{eqnarray}
  \label{e19}
 \left (\begin{array}{ccc}
0 & (1-p_1) & p_2e^{-\kappa}\\
p_1 & 0 &(1-p_2) \\
(1-p_1)e^\kappa & p_1 & 0 \end{array}\right )
\end{eqnarray}
For a random choice of the game we have $p_1=(h+s)/2,p_2=(h+r)/2$.
Then Eqs.~(12), (13) provide an exact solution.

The $M=3$ game with zero probability for holding the capital at the current
value, has peculiar properties: the probability distribution is nonzero for
odd differences in the capital after an odd number of time steps, and for
even differences after even time steps.

We checked that there are smooth limiting distributions for the
even and odd $n$'s, see Fig.~2.

 \begin{figure}
\large \unitlength=0.1in
\begin{picture}(42,12)
\put(-0.5,0.5){\includegraphics{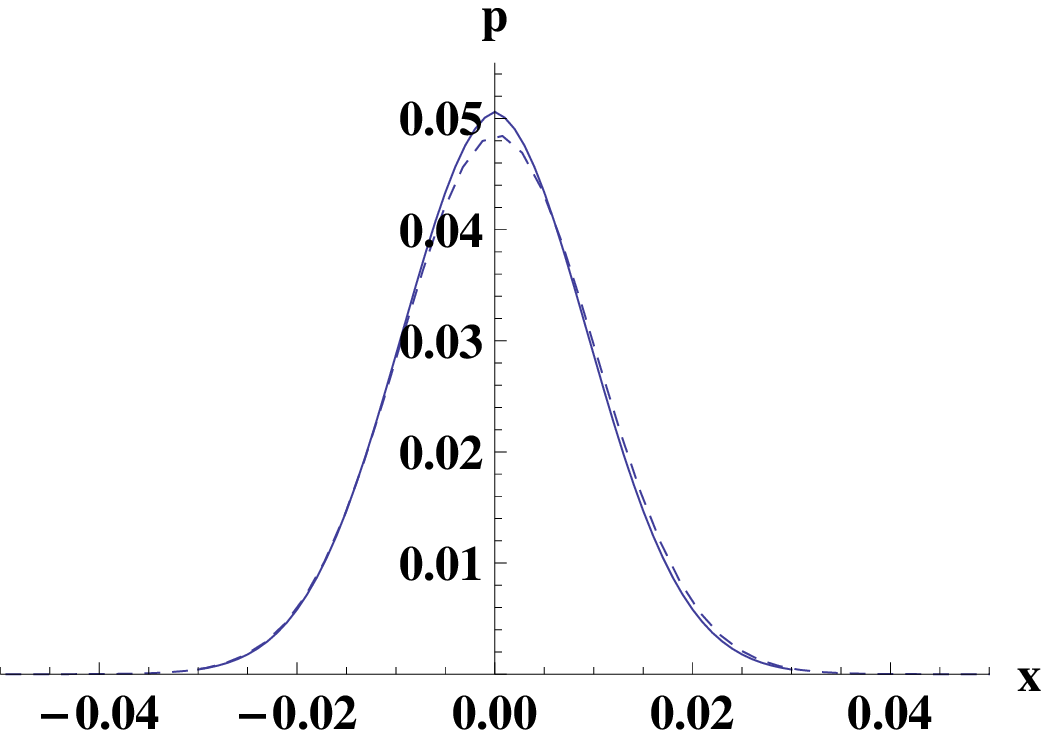}}
\put(16.5,0.5){\includegraphics{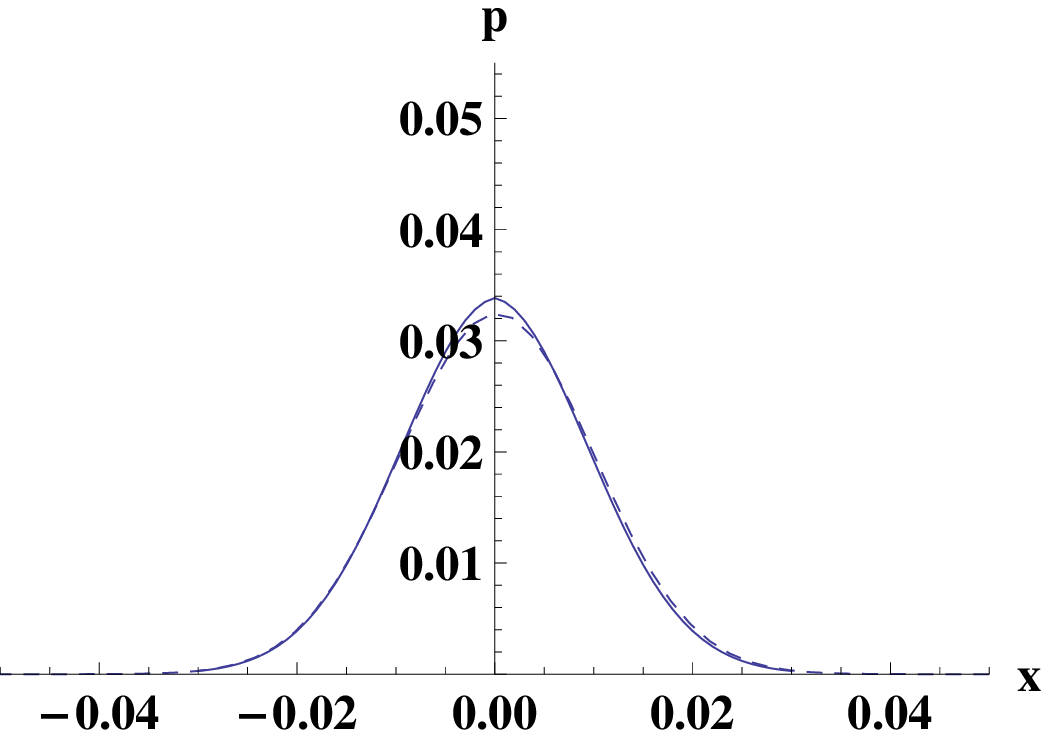}}
\put(7.7,0){\small{a.}} \put(26,0){\small{b.}}
\end{picture}
\caption{a) $p(x)=\hat P(n,t), x=n/t-r, t=1000$ for even $l$ for
the $M=3$ Parrondo's game with the same parameters as for Fig.~1.
b) $p(x)=\hat P(n,t)$ for odd $l$. The smooth lines are derived
according to our asymptotic formulas and Eq. (15), the dashed
lines correspond to the numerics.}
\end{figure}

It is important to find the possibility of transition between
different subphases.


Consider the $M=2$ game. For this game we have the right jump probabilities
$\barp_1,\barp_2$ and left jump probabilities $\barq_1,\barq_2$. For a second
game on a single axis we have the corresponding probabilities $h, 1-h$.

We obtain for the matrix
$\hat Q(\kappa)$
 \begin{eqnarray}
  \label{e18}
 \left (\begin{array}{cc}
 1-(p_1+q_1) &q_2+ p_2e^{-\kappa}\\
p_1+q_1e^{\kappa} &   1-(p_2+q_2)
\end{array}\right )
\end{eqnarray}
We give the characteristic equation to define the function $V(k)$
in the appendix.



Consider the case of random walks with memory. We denote the up
motion as +, down as -. Then the parameters of the motion depend on
$\alpha_1,\alpha_2$. We define the current state as
$X,\alpha_1,\alpha_2$. Then we get
 \begin{eqnarray}
 \label{e19}
P(X,+,\alpha,t+1)&=&P(X-1,\alpha,\beta,t)p_{\alpha,\beta}\nonumber\\
P(X,-,\alpha,t+1)&=&P(X+1,\alpha,\beta,t)(1-p_{\alpha,\beta})
 \end{eqnarray}
Let us introduce $w(X,t),y(X,t),z(X,t),h(X,t)$ for
$(-,-),(-,+),(+,-),(+,+)$ cases, with corresponding probabilities
of the right jumps $p_1,p_2,p_3,p_4$.

Then we have the master equations
 \begin{eqnarray}
 \label{e20}
w(X,t+1)&=&w(X+1,t)(1-p_1)+z(X+1,t)(1-p_3)\nonumber\\
y(X,t+1)&=&w(X-1,t)p_1+z(X-1,t)p_3\nonumber\\
z(X,t+1)&=&y(X+1,t)(1-p_2)+h(X+1,t)(1-p_4)\nonumber\\
h(X,t+1)&=&y(X-1,t)p_2+h(X-1,t)p_4\nonumber\\
 \end{eqnarray}

Performing a Fourier transformation we get
 \begin{eqnarray}
 \label{e21}
w(X,t)=v_{1}\exp[tu(X/t)],y(X,t)=v_{2}\exp[tu(X/t)]\nonumber\\
z(X,t)=v_{3}\exp[tu(X/t)],h(X,t)=v_{4}\exp[tu(X/t)]
 \end{eqnarray}
Finally,  we obtain a system of equations
 \begin{eqnarray}
 \label{e22}
v_1\lambda&=&v_1(1-p_1)e^{\kappa}+v_3(1-p_3)e^{\kappa}\nonumber\\
v_2\lambda&=&v_1p_1e^{-\kappa}+v_3p_3e^{-\kappa}\nonumber\\
v_3\lambda&=&v_2(1-p_2)e^{\kappa}+v_4(1-p_4)e^{\kappa}\nonumber\\
v_4\lambda&=&v_2p_2e^{-\kappa}+v_4p_4e^{-\kappa}
 \end{eqnarray}

In conclusion, we considered the general version of Parrondo's
games. The Parrondo's games, discovered 2 decades ago, describe a
counter-intuitive phenomenon in financial mathematics. Later the
model found many inter-disciplinary applications, when a random
mixture of two bad strategies gives a good strategy, also
conversely too many good witnesses result in low confidence. For
most applications it is critical to find the capital growth rate,
to find the variance of the distribution. We calculated not only
these characteristics of the models, also we found an exact
distribution function. The tail of the distribution is interesting
for the applications. We calculated also the asymptotics of the
distribution for large $t$. The u(x) function satisfies a highly
non-linear differential equation, but has an explicit expression
as the Legendre transform of an explicitly known ($M=1$)
resp.~more or less explicitly known function $(M>1)$ of the
momentum. For $M>1$ we have to do an eigenvalue analysis.
 An interesting finding is the existence of
oscillations in the probability distribution of the capital after
some rounds of gambling and the existence of two limiting
distributions. This is a typical situation with the real data of
stock fluctuations in financial market, and it is nice that our
toy model describes the phenomenon. We see also that different
sub-phases are possible while looking the large deviations from
the mean values, i.e.~the case in optimal coding \cite{gal}. How
realistic is the Parrondo's phenomenon? As we underlined, it
assumes a possibility to use a simple switch (degree of the mixing
between several strategies), either strengthening the system or
attenuating it. As we discussed in \cite{sa05}, the latter
situation with the possibility of anti-resonance, is typical for
the complex enough living systems.

%
%

We thank Bernard Derrida for discussions.  DBS thanks MOST
104-2811-M-001-102. grant, as well as Academia Sinica for the
support.

\end{document}